\documentclass{article}
\usepackage{FPSpro}

\begin{document}

\title{SPIN-DEPENDENT STRUCTURE FUNCTION $g_1$ AT SMALL $x$}

\author{B. I.~Ermolaev\\
{\em Ioffe Physico-Technical Institute, 194021 St.Petersburg, Russia}\\
M.~Greco\\
{\em Dep.t of Physics and INFN, University Rome III, Rome, Italy}\\
S. I.~Troyan\\
{\em St.Petersburg Institute of Nuclear Physics, 188300 Gatchina, Russia}
}

\maketitle

\baselineskip=11.6pt

\begin{abstract}
Accounting for double-logarithms of $x$ and running QCD coupling leads to
expressions for both the non-singlet and singlet components
of $g_1$. These expressions
manifest the Regge asymptotics when $x \to 0$ and differ considerably
from the DGLAP expressions at small values of $x$.
\end{abstract}

\newpage

\section{Introduction}

As is well known, deep inelastic scattering (DIS)
is one of the basic processes for probing
the structure of hadrons. From the theoretical point of view,
the inclusive cross section of DIS is a convolution of the leptonic and
hadronic tensors, with the information about
the structure of the hadrons
participating into DIS coming from the hadronic
tensor .
the forward Compton amplitude, when a deeply off-shell photon
with virtuality $q^2$ scatters off an on-shell hadron with momentum $p$.
The spin-dependent part, $W_{\mu \nu}^{spin}$, of the hadronic tensor
is parametrized in terms of two structure functions, $g_1$ and
$g_2$, as

\begin{equation}
\label{w}
W_{\mu \nu}^{spin} = \imath \epsilon_{\mu\nu\lambda\rho}
\frac{q_{\lambda}m}{pq}\Big[S_{\rho} g_1 +
\big(S_{\rho} - \frac{(Sq)}{pq} p_{\rho} \big) g_2 \Big]
\approx \imath\epsilon_{\mu\nu\lambda\rho}
\frac{q_{\lambda} m}{pq} \Big[ S_{\rho}^{||} g_1 +
S_{\rho}^{\perp}\big(g_1 + g_2 \big) \Big],
\end{equation}
so that $g_1$ is related to the longitudinal hadron spin-flip
scattering, whereas the sum
$g_1 + g_2$ is relative to the transverse spin-flips. In Eq.~(\ref{w}),
$m$ stands
for the hadron mass,
$S_{\rho}^{||}$ and  $S_{\rho}^{\perp}$ are the longitudinal and
transverse (with respect to the plane formed by $p$ and $q$)
components of the hadron spin  $S_{\rho}$.
Both $g_1$ and $g_2$
depend on $x = - q^2/2pq,~ 0< x \leq 1 $ and $Q^2 = - q^2 > 0$.
Obviously, small $x$ corresponds to $s = (p+q)^2 \approx 2pq \gg Q^2$.
In this case,
$S_{\rho}^{||} \approx  p_{\rho}/m$  and therefore the part of
$W_{\mu \nu}^{spin}$ related to $g_1$ does not depend on $m$. When
$Q^2 \gg m^2$, one can assume the factorization and regard
$W_{\mu \nu}^{spin}$
as a convolution of two objects (see Fig.~1).
\begin{figure}[t]
  \vspace{9.0cm}
  \includegraphics{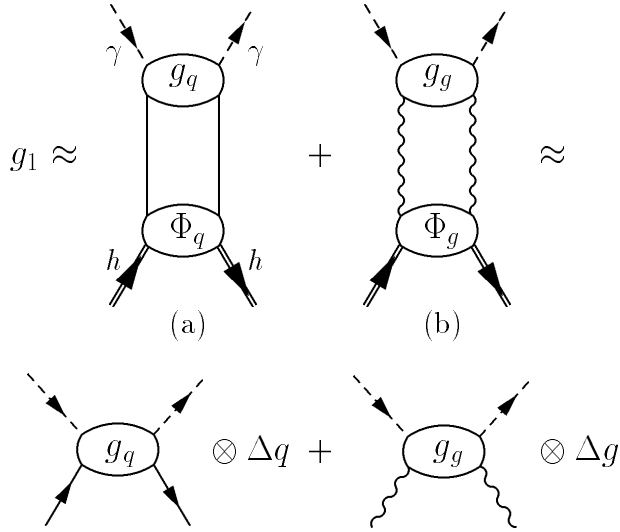}
  \caption{\it
    Representation of the hadronic $g_1$ as the convolution of
    the partonic $g_1$ and the initial parton densities.
    \label{fig1} }
\end{figure}
The first one is
the probability $\Phi$ ($\Phi = \Phi_q$ in Fig.~1a and $ \Phi = \Phi_g$ in
Fig.~1b) to find a polarized parton (a quark or a gluon) within the
hadron. The second one is the partonic tensor
$\widetilde{W}_{\mu \nu}^{spin}$ defined and parametrized
similarly to $W_{\mu \nu}^{spin}$.
Whereas
the partonic tensor $\widetilde{W}_{\mu \nu}^{spin}$,
i.e. the partonic structure functions $g_1$ and $g_2$, can
be studied within perturbative QCD,
$\Phi_{q, g}$ are essentially non-perturbative objects,
The lack of knowledge of  $\Phi$ is usually compensated by introducing
initial parton distributions which can be found from phenomenological
considerations. In doing so, the non-singlet component $g_1^{NS}$ of $g_1$
is usually expressed as a convolution of a piece
which we denote $g_q^{NS}$  from purely evolution, and the initial
polarized quark density
$\Delta q$:

\begin{equation}
\label{g1ns}
g_1^{NS} = g_q^{NS} \otimes \Delta q ~.
\end{equation}
Similarly, the singlet structure function $g_1^S$ is expressed in terms of
the evolution pieces $g_q$ and $g_g$ and the densities
of the polarized quarks and gluons, $\Delta q$ and $\Delta g$:

\begin{equation}
\label{g1s}
g_1^S = g_q \otimes \Delta q + g_g \otimes \Delta g~.
\end{equation}
The subscripts $q,~g$ in Eqs.~(\ref{g1ns},\ref{g1s}) refer to the kind of
the initial partons, as is shown at Fig.~1.
We remind that there is no rigorous procedure for calculating
$\Delta q$ and $\Delta g$. They have to be found from phenomenological
considerations.
On the contrary, there are regular perturbative
methods for calculating the evolution parts $g_q^{NS}$ and  $g_{q,g}$
of $g_1$. When  $Q^2$ is much greater than the starting point  $\mu^2$
of the $Q^2$-evolution and at the same time $x \ll 1$, it is convenient to
rewrite  Eqs.~(\ref{g1ns},\ref{g1s}) in
the form of the Mellin integral:

\begin{eqnarray}
\label{gmellin}
&&g_1^{NS} = \int_{- \imath \infty}^{\imath \infty}
\frac{d\omega}{2\pi\imath} (1/x)^{\omega}
C_{NS}(\omega)e^{\gamma_{NS}(\omega) \ln(Q^2/\mu^2)} ~, \\  \nonumber
&&g_1^S =
\int_{- \imath \infty}^{\imath \infty}
\frac{d\omega}{2\pi\imath} (1/x)^{\omega}
\Big[ C_q(\omega) \Delta q(\omega) +
C_g(\omega) \Delta g(\omega) \Big]
e^{\gamma_S(\omega) \ln(Q^2/\mu^2)} ~,
\end{eqnarray}
with $C_{NS}(\omega), C_{q,g}(\omega)$ being the coefficient functions and
$\gamma_{NS}(\omega),\gamma_S(\omega)$ the non-singlet and singlet
anomalous dimensions respectively. The anomalous dimensions control the
$Q^2$-evolution and the coefficient functions govern the $s$-evolution
which, at fixed $Q^2$, is equivalent to the $x$-evolution.

The best known instrument to calculate the DIS structure functions
is the DGLAP\cite{dglap} approach. In the DGLAP framework, both
the coefficient functions and the anomalous dimensions are
perturbatively known and
represented by their one-loop, (or leading order (LO)) \cite{dglap}
and two-loop
(next-to-leading order (NLO))contributions. The relevant
Ref.s concerning the NLO
calculations can be found in the review \cite{neerven}.
The remaining ingredients to the rhs of Eq.~(\ref{gmellin}),
$\Delta q$ and $\Delta g$ can be taken, for example, from Ref.~\cite{a}.
DGLAP provides a quite good description
of the
experimental data\cite{a}.
The extrapolation of DGLAP into
the small-$x$ region predicts an asymptotic behavior
$\sim \exp(\sqrt{C \ln(1/x)\ln\ln Q^2})$ for all DIS
structure functions (with different factors $C$). However, from a
theoretical point of view, such  an extrapolation at
the small-$x$ is
rather doubtful. In particular, it neglects in a systematical way
contributions ot the type $\sim (\alpha_s \ln^2(1/x))^k$ which are small
when $x \sim 1$ but become relevant when $x \ll 1$.
The total resummation of
these double-logarithmic (DL) contributions was made in
Refs.~\cite{ber1} and Ref.~\cite{ber} for the non-singlet $(g_1^{NS})$ and
singlet $g_1$ respectively, and it  leads to the Regge (power-like)
asymptotics
$g_1 (g_1^{NS}) \sim (1/x)^{\Delta^{DL}} ((1/x)^{\Delta^{DL}_{NS}})$,
with $\Delta^{DL}, \Delta^{DL}_{NS}$ being the
intercepts calculated ib the double-logarithmic approximation (DLA).
The weak point of this resummation in Refs.~\cite{ber1},\cite{ber} is the
assumption that $\alpha_s$  is kept fixed (at some unknown scale). It
leads therefore to the value of
the intercepts $\Delta^{DL}, \Delta^{DL}_{NS}$  explicitly
depending on this unknown coupling,
while  $\alpha_s$  is well-known to be running. The results of
Refs.~\cite{ber1},\cite{ber} had led many authors
(see e.g.\cite{kw}) to suggest that the DGLAP parametrization
$\alpha_s = \alpha_s(Q^2)$ has to be used. However, according to results
of  Ref.~\cite{egt1}, such a parametrization is indeed
correct for $x \sim 1$ only and
cannot be used for $x \ll 1$. The appropriate dependence of  $\alpha_s$
suggested in Ref.~\cite{egt1}, has been successfully used to calculate
both $g_1^{NS}$
and $g_1$ singlet at small $x$ in Refs.~\cite{egt2}.
In the present talk we review these results.

Instead of a direct study of $g_1$ like it is done in DGLAP,
it is more convenient to consider the
forward Compton amplitude $A$ for the photon-parton scattering
related to $g_1$, with $\Delta q =1,~ \Delta g = 0$, as follows:

\begin{equation}
\label{g1a}
g_1(x, Q^2) = \frac{1}{\pi} \Im_s A(s, Q^2) ~.
\end{equation}

As already stated above, we cannot use DGLAP for studying $g_1$ or $A$ at
small $x$ because it does not
not include the total resummation of the double- and single-logarithms of
$x$ for the anomalous dimensions and coefficient functions, and also
the $\alpha_s$-parametrization used is valid for $x$ not far from 1.

Then, in order to
account for  the double-logs of both $x$ and  $Q^2$,
we need to construct a kind of two-dimensional
evolution equations that would combine both the $x$- and $Q^2$-
evolutions.

Such equations should sum up the contributions of
the Feynman graphs involved to all orders in $\alpha_s$. Some of those
graphs have either ultraviolet or infrared (IR) divergences. The
ultraviolet divergences are regulated by the usual renormalization
procedure. In order  to regulate the IR ones, we introduce an IR
cut-off $\mu$ in the transverse momentum space for the
momenta $k_i$ of all virtual quarks and gluons:

\begin{equation}
\label{mu}
\mu < k_{i \perp}
\end{equation}
where $k_{i \perp}$ stands for the transverse (with respect to the plane
formed by the external momenta $p$ and $q$) component of $k_i$. This
technique
of regulating the IR divergences was suggested by Lipatov and used
first in Ref.~\cite{kl} for quark-quark scattering. Using this cut-off
$\mu$, $A$ acquires a dependence on $\mu$. Then, one can
evolve $A$ with respect to $\mu$, constructing the appropriate Infrared
Evolution Equations (IREE). As $A = A(s/\mu^2, Q^2/\mu^2)$,

\begin{equation}
\label{lhs}
- \mu^2 \partial A / \partial \mu^2 =
 \partial A/ \partial \rho + \partial A/ \partial y
\end{equation}
where $\rho = \ln(s/\mu^2)$ and $y = \ln(Q^2/\mu^2)$. Eq.~(\ref{lhs})
represents the lhs of the IREE for A.
In order to write the rhs of the IREE, it is convenient to use the
Sommerfeld-Watson transform
\begin{equation}
\label{mellin}
 A(s, Q^2) =
\int_{- \imath \infty}^{\imath \infty}\frac{d \omega}{2 \pi \imath}
(s/ \mu^2)^{\omega}
\xi(\omega)  F(\omega, Q^2)
\end{equation}
where $\xi(\omega)$ is the negative signature factor,
$\xi(\omega) =  [1 - e^{ - \imath \pi \omega}]/ 2 \approx
 \imath \pi \omega / 2$.
It must be noted that  the transform inverse to Eq.~(\ref{mellin})
involves the imaginary parts of $A$:

\begin{equation}
\label{invmellin}
F(\omega, Q^2) =  \frac{2}{\pi \omega}\int_0^{\infty} d \rho
e^{- \rho \omega}
\Im A(s, Q^2) ~.
\end{equation}
Notice that, contrary to the amplitude $A$, the structure function $g_1$
does not have any signature and therefore  $\xi(\omega) = 1$ when
the transform ~(\ref{mellin}) is applied directly to $g_1$.

\section{infrared evolution equations for $g_1$ }

When the factorization depicted in Fig.~1 is assumed, the calculation
of $g_1$ (for semplicity
we will omit the superscript ``s''
for$g_1$ singlet, though we use the notation $g_1^{NS}$ for the
non-singlet $g_1$ ) is
reduced to calculating the Feynman graphs contributing  the partonic
tensor $\tilde{W}_{\mu \nu}^s$ depicted as the upper blobs
in Fig.~1.
Both cases, (a), when the virtual photon scatters off the
nearly on-shell polarized quark, and (b), when the quark is replaced by
the polarized gluon, should be taken into account. Therefore, in contrast
to Eq.~(\ref{g1a}), we need to introduce two Compton amplitudes: $A_q$
and $A_g$ corresponding to the upper blob in Fig.~1a and Fig.~1b
respectively. The subscripts ``q'' and ``g''
refer to the initial partons. Therefore,

\begin{equation}
\label{g1qg}
g_1(x, Q^2) = g_q(x, Q^2) + g_q(x, Q^2),
\end{equation}
where
\begin{equation}
\label{gqg}
~g_q = \frac{1}{\pi} \Im_s A_q(s, Q^2),
~g_g = \frac{1}{\pi} \Im_s A_g(s, Q^2) ~
\end{equation}

Let us now construct the IREE for
the amplitudes $A_{q,g}$ related to $g_1$.
To this aim, let us consider a virtual parton
with minimal $k_{\perp}$. We call such a parton the softest one.
If it is a gluon, its DL contribution can be
factorized, i.e. its DL contribution comes from the graphs where
its propagator is attached to the external lines. As the gluon propagator
cannot be attached to photons, this case is absent in IREE for $A_{q,g}$.
The second option is when the softest partons are a $t$-channel
quark-antiquark or gluon pair. It leads us to the IREE depicted in Fig.~2.
\begin{figure}[t]
  \vspace{9.0cm}
  \includegraphics{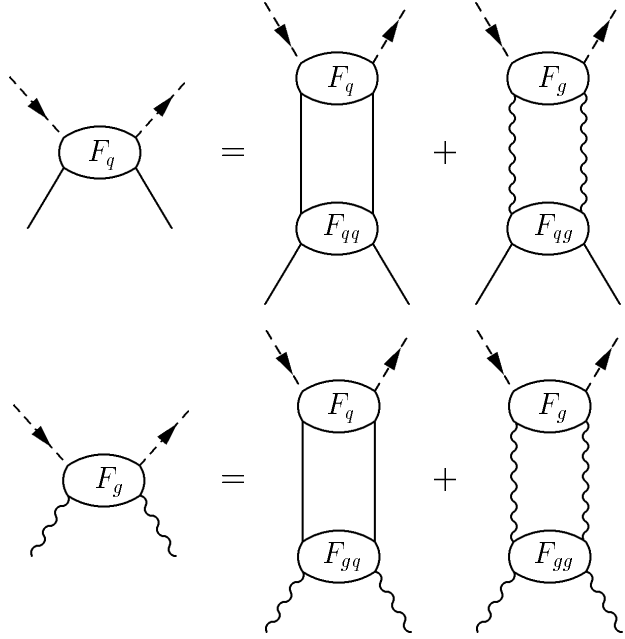}
  \caption{\it
              Infrared evolution equations for the amplitudes $A_q, A_g$.
    \label{fig2} }
\end{figure}
Applying the operator $- \mu^2 \partial / \partial \mu^2$ to it,
combining the result with Eq.~(\ref{lhs}) and
using (\ref{mellin}), we arrive at the following system of equations:

\begin{eqnarray}
\label{system}
\big( \omega + \frac{\partial}{\partial y}\big) F_q(\omega, y) &=&
\frac{1}{8 \pi^2} \big[ F_{qq}(\omega) F_q(\omega, y)  +
F_{qg}(\omega) F_g (\omega, y)\big] ~, \nonumber \\
\big( \omega + \frac{\partial}{\partial y}\big) F_g(\omega, y) &=&
\frac{1}{8 \pi^2} \big[ F_{gq}(\omega) F_q(\omega, y)  + F_{gg}(\omega)
 F_g(\omega, y) \big] ~.
\end{eqnarray}
The amplitudes
$F_q, F_g$ are related to  $A_q, A_g$ through the transform
(\ref{mellin}).
The Mellin amplitudes $F_{ik}$, with $i,k = q,g$, describe the
parton-parton forward scattering. They contain DL contributions to
all orders in $\alpha_s$. We can introduce the new anomalous
dimensions $H_{ik} = (1/8\pi^2)F_{ik}$. The
subscripts ``q,g'' correspond to
the DGLAP-notation.
Solving this system of eqs. and using Eq.~(\ref{gqg}) leads to

\begin{eqnarray}
\label{gqgg}
g_q(x, Q^2) &=& \int_{- \imath \infty}^{\imath \infty}
\frac{d \omega}{2 \pi \imath} (1/ x)^{\omega}
\Big[C_+(\omega) e^{\Omega_+ y} +
C_-(\omega) e^{\Omega_- y} \Big] ~, \\ \nonumber
g_g(x, Q^2) &=& \int_{- \imath \infty}^{\imath \infty}
\frac{d \omega}{2 \pi \imath} (1/ x)^{\omega}
\Big[C_+(\omega) \frac{X + \sqrt{R}}{2H_{qg}} e^{\Omega_+ y} +
C_-(\omega) \frac{X - \sqrt{R}}{2H_{qg}} e^{\Omega_- y} \Big]~.
\end{eqnarray}

The unknown factors $C_{\pm}(\omega)$
have to be specified and will be discussed later. All other factors in
Eq.~(\ref{gqgg}) can be
expressed in therms of $H_{ik}$:

\begin{eqnarray}
\label{xromega}
X = H_{gg} - H_{qq}~,\qquad R=(H_{gg}-H_{qq})^2+4H_{qg}H_{gq}~,\\ \nonumber
\Omega_{\pm} =\frac{1}{2}\left[H_{qq}+H_{gg}\pm
\sqrt{(H_{qq} - H_{gg})^2 + 4H_{qg}H_{gq}} \right]~.
\end{eqnarray}

The anomalous dimension matrix $H_{ik}$ was calculated in Ref.~\cite{egt2}:

\begin{eqnarray}
\label{solh}
H_{gg}&=&\frac{1}{2} \Big(\omega+Y+\frac{b_{qq}-b_{gg}}{Y}\Big)~,\quad
H_{qq}=\frac{1}{2} \Big(\omega + Y - \frac{ b_{qq} - b_{gg}}{Y}\Big),
\\ \nonumber
H_{gq} &=& -\frac{b_{gq}}{Y}, ~~~~~~~~H_{qg} = -\frac{b_{qg}}{Y} ~.
\end{eqnarray}
where

\begin{equation}
\label{Y}
Y =\!-\sqrt{\!\Big(\omega^2\!-\!2(b_{qq}\!+\!b_{gg})\!+\!
\sqrt{\!\left[(\omega^2\!-\!2(b_{qq}\!+\!b_{gg}))^2\!-\!4(b_{qq}\!-\!b_{gg})^2
\!-\!16b_{qg}b_{gq}\!\right]}\Big)/2}~,
\end{equation}
\begin{equation}
\label{b}
b_{ik} = a_{ik} + V_{ik} ,
\end{equation}
\begin{equation}
\label{aik}
a_{qq} = \frac{A(\omega)C_F}{2\pi},
~a_{gg} = \frac{2A(\omega) N}{\pi},~
a_{gq} = -\frac{n_f A'(\omega)}{2\pi},~
a_{qg} =\frac{ A'(\omega)C_F}{\pi} ~,
\end{equation}
and
\begin{equation}
\label{vborn}
V_{ik} =\frac{ m_{ik}}{\pi^2}D(\omega) ,
\end{equation}
with
\begin{equation}
\label{m}
m_{qq} =  \frac{C_F}{2N},~
m_{gg} = -2N^2,~
m_{qg} = n_f \frac{N}{2},~
m_{gq} = -N C_F.
\end{equation}
We have used here the notations $C_F = 4/3, N = 3$ and $n_f = 4$.
The quantities $A(\omega)$ and $D(\omega)$ account for the running of
$\alpha_s$. They are given by the following
expressions:

\begin{equation}
\label{a}
A(\omega) = \frac{1}{b} \Big[\frac{\eta}{\eta^2 + \pi^2} -
\int_{0}^{\infty} \frac{d \rho e^{- \omega \rho}}
{(\rho + \eta)^2 + \pi^2}\Big] ~,
\end{equation}
\begin{equation}
\label{d}
D(\omega) = \frac{1}{2b^2} \int_0^{\infty} d \rho e^{- \omega \rho}
\ln \big( (\rho + \eta)/ \eta\big)
\Big[  \frac{\rho + \eta}{(\rho + \eta)^2 + \pi^2} +
\frac{1}{\rho + \eta}\Big]
\end{equation}
with $\eta = \ln(\mu^2/ \Lambda_{QCD}^2)$ and $b = (33 - 2 n_f)/ 12 \pi$.
$A'$ is defined as  $A$ with the $\pi^2$ term dropped out.

Finally we have to specify the coefficients functions $C_{\pm}$ appearing
in Eq.~(\ref{gqgg}). When
$Q^2 = \mu^2$,
 \begin{equation}
\label{match}
g_q = \tilde{\Delta} q(x_0), ~~~~~~~~g_g = \tilde{\Delta} g(x_0)
\end{equation}
where $ \tilde{\Delta} q(x_0)$ and $ \tilde{\Delta} g(x_0)$ are
the input distributions of the polarized partons at $x_0 = \mu^2/s$.
They do not depend on $Q^2$.
Eq.~(\ref{match}) allows us to express $C_{\pm}(\omega)$ in terms of
 $\Delta q(\omega)$ and $\Delta g(\omega)$,
which are related to $\tilde{\Delta} q(x_0)$
and $\tilde{\Delta} g(x_0)$ through the ordinary Mellin transform. Indeed,

\begin{equation}
\label{matchomega}
C_+ + C_- = \Delta q,
~~~~C_+\frac{X + \sqrt{R}}{2H_{qg}} +
C_- \frac{X - \sqrt{R}}{2H_{qg}} = \Delta g ~ ,
\end{equation}
with both $\Delta q$ and $\Delta g$ depending on $\omega$. This leads to the
following expressions for $g_q$ and $g_g$:
\begin{equation}
\label{gqsol}
g_q(x, Q^2) = \int_{- \imath \infty}^{\imath \infty}
\frac{d \omega}{2 \pi \imath} (1/ x)^{\omega}
\Big[ \Big(  A^{(-)} \Delta q +
B \Delta g \Big)  e^{\Omega_+ y} +
 \Big( A^{(+)} \Delta q - B \Delta g \big) e^{\Omega_- y} \big]~,
\end{equation}

\begin{equation}
\label{ggsol}
g_g(x, Q^2) = \int_{- \imath \infty}^{\imath \infty}
\frac{d \omega}{2 \pi \imath} (1/ x)^{\omega}
\Big[\Big( E \Delta q + A^{(+)}\Delta g
\Big) e^{\Omega_+ y} +
\Big(- E\Delta q +  A^{(-)}\Delta g \Big)
 e^{\Omega_- y}\Big]
\end{equation}
with
\begin{equation}
\label{abe}
A^{(\pm)} = \Big(\frac{1}{2} \pm \frac{X}{2 \sqrt{R}}\Big)~,\quad
B = \frac{H_{qg}}{\sqrt{R}}~,\quad
E = \frac{H_{gq}}{\sqrt{R}}~.
\end{equation}

Eqs.~(\ref{gqsol}, (\ref{ggsol}) express $g_1$ in terms of the
parton distributions $\Delta q(\omega)$ and $\Delta g(\omega)$, which
are related to the distributions
  $\tilde{\Delta} q(x_0)$ and $\tilde{\Delta} g(x_0)$ at very low $x$:
 $x_0 \approx \mu^2/s \ll 1$. Therefore, they hardly
can be found from experimental data. It is much more useful to
express $g_q, g_g$ in terms of the initial parton densities
 $\tilde{\delta} q$ and $\tilde{\delta} g$
defined at $x \sim 1$. We can do it, using the evolution of
  $\tilde{\Delta} q(x_0)$,~ $\tilde{\Delta} g(x_0)$ with respect to
$s$. Indeed, the $s$-evolution of $\tilde{\delta} q, \tilde{\delta} q$
from $s \approx \mu^2$ to $s \gg \mu^2$
 at fixed $Q^2$ ~$(Q^2 = \mu^2)$ is equivalent to their $x$-evolution
from $x \sim 1$ to $x \ll 1$. In the $\omega$-space, the system of IREE
for the parton distributions looks quite similar to  Eqs.~(\ref{system}).
However, the eqs for $\Delta q,~ \Delta g$ are now algebraic because they
do not depend on $Q^2$:

\begin{eqnarray}
\label{systeminput}
  \Delta q (\omega) &=& (<e^2_q>/2) \delta q(\omega) + (1 /\omega)
\left[H_{qq}(\omega) \Delta q(\omega)  +
H_{qg}(\omega) \Delta g (\omega) \right]~, \nonumber \\
 \Delta g(\omega) &=& (<e^2_q>/2)\hat{\delta g}(\omega) + (1/\omega)
\left[H_{gq}(\omega) \Delta q(\omega)  +
H_{gg}(\omega) \Delta g(\omega) \right] ~.
\end{eqnarray}
 where $<e^2_q>$ is the sum of the quark electric
charges ($<e^2_q> = 10/9$ for $n_f = 4$) and
$\hat{\delta g} \equiv  -  (A'(\omega)/\pi \omega^2) \delta g$
is the starting
point of the evolution of the gluon density $\delta g$. It corresponds to
Fig.~1b where the upper blob is substituted by the quark box.
Solving Eqs.~(\ref{systeminput}), we
obtain:

\begin{equation}
\label{inputq}
\Delta q=
\frac{(<e^2_q>/2)
\big[\omega (\omega -H_{gg}) \delta q + \omega H_{qg}\hat{\delta g}\big] }
{\big[\omega^2 - \omega(H_{qq} + H_{gg}) + (H_{qq}H_{gg} -
H_{qg}H_{gq})\big]}~,
\end{equation}
\begin{equation}
\label{inputg}
\Delta g=
 \frac{(<e^2_q >/2)
\big[\omega H_{gq} \delta q + \omega(\omega - H_{qq})\hat{\delta} g \big]}
{\big[\omega^2 - \omega(H_{qq} + H_{gg}) + (H_{qq}H_{gg} -
H_{qg}H_{gq})\big]}  ~.
\end{equation}

Then Eqs.~(\ref{gqsol},\ref{ggsol},\ref{inputq},\ref{inputg}) express
$g_1$ in
terms of the initial parton densities  $\delta q, \delta g$.

When we put $H_{qg}= H_{gq}= H_{gg} = 0$ and do not sum over $e_q$, we
arrive at the expression for the non-singlet structure function $g_1^{NS}$:
Obviously, in this case $A^{(+)} = B = E = \Omega_- = 0,~
A^{(-)} = 1,~~ \Omega_+ = H_{qq}$. However, the nonsinglet
anomalous dimension $ H_{qq}$ should be  calculated in the limit
 $b_{gg} = b_{qg} = b_{gq} = 0$. We denote such  $H_{qq} \equiv H^{NS}$.
The explicit expression for it is:
 \begin{equation}
\label{f0ns}
H^{NS} =
(1/2) \Big[\omega - \sqrt{\omega^2 - 4 b_{qq}} \Big]~.
\end{equation}
Therefore, we arrive at

 \begin{equation}
\label{gns}
g_1^{NS} = \frac{e^2_q}{2} \int_{ -\imath \infty}^{\imath \infty}
\frac{d \omega}{2 \pi \imath}
\Big(\frac{\omega \delta q}{\omega - H^{NS} } \Big)
\Big( 1/x \big)^{\omega}
\Big( Q^2/\mu^2\big)^{H^{NS}  } ~.
\end{equation}

\section{Small-$x$ asymptotics for $g_1$ }

Before making use of Eqs.~(\ref{gqsol}),(\ref{ggsol} and (\ref{gns}) for
calculating $g_1$ at small but finite values of $x$, let us discuss its
asymptotics.
When $x \to 0$ and $Q^2 \gg \mu^2$, one can
neglect contributions with $\Omega_-$ in  Eqs.~(\ref{gqgg}). As is
well known,
$g_1 \sim (1/x)^{\omega_0}$ at $x \to 0$, with $\omega_0$ being the position
of the leading singularity of the integrand of $g_1$ .
According to  Eqs.~(\ref{solh}),  the leading singularity, $\omega^{NS}$
for $g_1^{NS}$ is the rightmost
root of the equation
 \begin{equation}
\label{singns}
\omega^2 - 4 b_{qq} = 0
\end{equation}
while the leading singularity,
$\omega_0$ for $g_1$ is the rightmost root of

 \begin{equation}
\label{sings}
\omega^4 - 4 (b_{qq} + b_{gg})\omega^2 + 16
(b_{qq} b_{gg} -   b_{qg} b_{gq}) = 0~.
\end{equation}

In our approach, all factors $b_{ik}$ depend on
$\eta = \ln(\mu^2/\Lambda_{QCD})$, so the roots of
Eqs.~(\ref{singns},\ref{sings}) also depend on $\eta$.
This dependence is plotted in Fig.~3 for $\omega^{NS}$  and in Fig.~4
for $\omega_0$.
\begin{figure}[t]
  \vspace{5.cm}
  \includegraphics{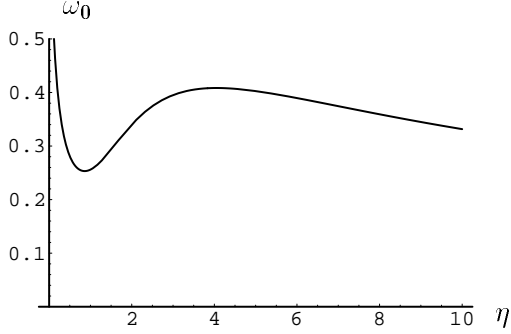}
  \caption{\it
    Dependence of  the leading singularity $\omega_0^{NS}$ on
    $\eta$.
    \label{fig3} }
\end{figure}
\begin{figure}[t]
  \vspace{5.0cm}
  \includegraphics{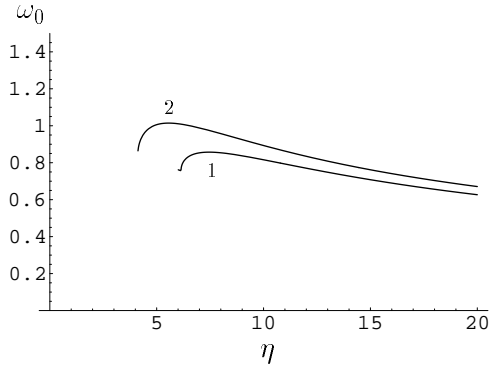}
  \caption{\it
    Dependence of the leading singularity $\omega_0$ on $\eta$.
    Curve~2 corresponds to the case where gluon contributions only
    are taken into account; curve~1 is the result of accounting for
    both gluon and quark contributions.
    \label{fig4} }
\end{figure}
Both the curve in Fig.~3 and the curve~1 in Fig.~4 have
a maximum. We denote this maximum as the intercept. Therefore,

\begin{eqnarray}
\label{g1asympt}
&&g_1^{NS}\sim e^2_q \delta q
(1/x)^{\Delta_{NS}}(Q^2/\mu^2)^{\Delta_{NS}/2}~,\\ \nonumber
&&g_1 \sim (<e^2_q>/2) [Z_1 \delta q + Z_2 \delta_g]
(1/x)^{\Delta_S}(Q^2/\mu^2)^{\Delta_S/2},
\end{eqnarray}
and we find for the intercepts
\begin{equation}
\label{intercepts}
\Delta_{NS} \approx 0.4,~\Delta_S \approx 0.86
\end{equation}
and
$Z_1 = - 1.2,~ Z_2 = -0.08$.
This implies that $g_1^{NS}$ is positive when $x \to 0$ whereas $g_1^S$
can be either positive or negative, depending on the relation between
$\delta q$ and $\delta g$. In particular, $g_1$ is positive when
\begin{equation}
\label{posit}
15\delta q + \delta g < 0 .
\end{equation}
otherwise it is negative. In other words, the sign of $g_1$ at small $x$
can be positive if the initial gluon density is negative and large.

\section{$g_1^{NS}$ at finite values of $x$}

Let us estimate the impact of the total resummation of DL and SL
contributions on $g_1^{NS}$.
According to Eq.~(\ref{gns}), the value of $g_1(x, Q^2)^{NS}$
depends both on the perturbative terms and on the inputs
$\Delta q$. The latter can be obtained by fitting the experimental data
and
it is known (see e.g. \cite{a}) that widely different formulae for
$\Delta q$ can be used. In order to avoid discussing the fitting procedure
and  as in this paper we present our results only the
perturbative part of $g_1^{NS}$, we can
assume
\begin{equation}
\label{quarkinput}
 \Delta q = \delta(1-x) ~.
\end{equation}

Then let us calculate the ratio

\begin{equation}
\label{ratio}
R = g_1^{NS}/ \tilde{g_1}^{NS}~,
\end{equation}
where $\tilde{g_1}^{NS}$ is the LO DGLAP non-singlet $g_1$. The results of
a numerical calculations for $R$ at $Q^2 = 20$GeV$^2$,
 $\mu = 1.5$GeV are shown in Table~1.
\begin{table}[t]
\begin{center}
\caption{\it Ratio $R$ of structure functions $g_1^{NS}$ calculated in DLA
and  DGLAP LO.}
\begin{tabular}{|c|c|}
\hline\hline
$x$ & $R$ \\
\hline\hline
0.1 & 1.60 \\
0.01 & 2.59 \\
0.001 & 4.33 \\
0.0001 & 7.46 \\
\hline\hline
\end{tabular}
\end{center}
\label{r}
\end{table}

It shows that the impact of the total resummation of DL contributions
is negligible for $x \geq 0.1$ but it grows fast with
decreasing of $x$, achieving the values
$R = 2.6$ at $x = 10^{-2}$ and $R = 7.5$ at $x = 10^{-4}$.

\section{Conclusion}
\label{CONCLUSION}

The total resummation of the most singular $(\sim \alpha^n_s/\omega^{2n +
1})$ terms in the expressions for the anomalous dimensions and the
coefficient
functions leads to
the expressions of
Eqs.~(\ref{g1qg},\ref{gqsol},\ref{ggsol},\ref{gns}) for the
singlet and the non-singlet structure functions $g_1$.  It
guarantees the Regge
(power-like) behavior (\ref{g1asympt}) of $g_1,~ g_1^{NS}$ when $x \to 0$,
with the intercepts given by  Eq.~(\ref{intercepts}). The intercepts
$\Delta_{NS}, \Delta_S$ are  obtained with the running QCD
coupling effects taken into account. The value of
the non-singlet intercept $\Delta_{NS} \approx 0.4$ is now confirmed by several
independent analysis
\cite{kat} of experimental data and our result $\Delta_S \approx 0.86$ is
in a good agreement with the estimate of Ref.~\cite{koch}:
$\Delta_S = 0.88 \pm 0.14$ obtained from analysis of the HERMES data.
Eq.~(\ref{gns}) states
that $g_1^{NS}$ is positive both at $x \sim 1$ and at $ x \ll 1$.
The situation concerning the singlet $g_1$ is more involved: being
positive at
$x \sim 1$, the singlet $g_1$ can remain positive at $x \ll 1$
only if the initial
parton densities obey Eq.~(\ref{posit}), otherwise it becomes negative.
The ratio of our results versus the DGLAP ones for non-singlet $g_1$ is
given in Table~1. It shows explicitly that the impact of high-order
DL contributions is small at $x \geq 0.1$ but it grows fast when $x$ is
approaching   $10^{-3}-10^{-4}$.

\section{Acknowledgment}

The work is supported by grant RSGSS-1124.2003.2

\end{document}